\documentclass[english]{paper}
\usepackage[T1]{fontenc}
\usepackage[latin9]{inputenc}
\usepackage[a4paper]{geometry}
\usepackage{bm}
\usepackage{amsmath}
\usepackage{amssymb}
\usepackage{graphicx}

\makeatletter
\newcommand{\lyxaddress}[1]{
\par {\raggedright #1
\vspace{1.4em}
\noindent\par}
}

\makeatother

\usepackage{babel}
\begin{document}

\title{Tetramers of two heavy and two light bosons}

\author{Pascal Naidon}
\maketitle

\lyxaddress{RIKEN Nishina Centre, RIKEN, Wak{\=o}, 351-0198 Japan}
\begin{abstract}
This article considers the bound states of two heavy and two light
bosons, when a short-range force attracts the bosons of different
mass, and a short-range force repel the light bosons. The existence
of such four-body bound states results from the competition between
these two forces. For a given strength of the attraction, the critical
strength of the repulsion necessary to unbind the four particles is
calculated. This study is motivated by the experimental realisation
of impurity atoms immersed in an atomic Bose-Einstein condensate,
and aims at determining in which regime only one boson contributes
to binding two impurities.
\end{abstract}

\section{Introduction}

Since the observation of Efimov trimers in ultra-cold atomic gases~\cite{Kraemer2006,Braaten2007,Ferlaino2010,Naidon2017},
there have been several theoretical~\cite{Platter2004,Yamashita2006,Hammer2007,Stecher2009,Hadizadeh2011,Deltuva2012b,Deltuva2012c,Deltuva2013}
and experimental~\cite{Ferlaino2009,Pollack2009,Dyke2013} studies
looking into the four-body bound states that exist around Efimov trimers.
These tetramers are not ``Efimov tetramers'', in the sense that they
do not result from a four-body Efimov effect~\cite{Castin2010},
yet they possess a universal character that originates, like the Efimov
trimers, from the three-body Efimov effect. In particular, they follow
the three-body Efimov scaling, such that each Efimov trimer is accompanied
by one or several tetramer bound states or resonances.

The case of heteronuclear systems is particularly interesting, because
the Efimov effect is enhanced when two heavy particles are attracted
to a light particle. In this case, the light particle may be seen
as a glue that binds the two heavy particles. This picture is clearly
seen in the limit of large mass ratio between the heavy and light,
where the Born-Oppenheimer approximation can be applied~\cite{Naidon2017}.
In this approximation, a light particle of mass $m$ creates an Efimov
attraction potental $V(R)=-\frac{\hbar^{2}}{MR^{2}}(\vert s_{0}\vert^{2}-1/4)$
between two heavy particles of mass $M$ separated by $R$, where
\begin{equation}
\vert s_{0}\vert^{2}-1/4=\frac{M}{2m}(0.567143)^{2},\label{eq:s0}
\end{equation}
resulting in a discrete-scale-invariant spectrum with a discrete scaling
factor $e^{\pi/\vert s_{0}\vert}$.  It turns out that a light particle
can also glue three heavy particles together. The studies on the heteronuclear
tetramers have so far focused on this situation when the three heavy
particles are bosons~\cite{Wang2012b,Blume2014b,Schmickler2016,Schmickler2017}
or fermions~\cite{Blume2010a,Blume2010,Castin2010,Blume2012,Bazak2017}. 

Another interesting possibility is the case of two heavy particles
and two light particles. In this case, in the large mass ratio limit,
the two light particles are expected to provide twice the Efimov attraction
of a single light particle, resulting in an Efimov series of tetramers
that scale with a smaller scaling factor than that of the Efimov trimers.
However, this is expected only when the interaction between the light
particles may be neglected. If, on the other hand, the two light particles
strongly repel each other, it might not be possible for the two light
particles to bind to the two heavy particles; in other words, trimers
are possible, while tetramers may not exist. The purpose of this article
is to study how the spectrum changes between these two limiting cases,
and determine the critical strength of repulsion required to unbind
tetramers for a given attraction between the light and heavy particles.

This study is motivated by the question of heavy impurities immersed
in a condensate of light bosons. The general treatment of this problem
requires to include many excitations of the condensate created by
the heavy impurities. However, when the light bosons repel each other
sufficiently, a single excitation may be enough to capture the polaron
physics of two impurities~\cite{Naidon2016b}. The present study
aims at quantifying the strength of repulsion necessary to reach this
regime. It should be noted that in an ultra-cold atomic experiment,
the light bosons actually experience an attractive interatomic force.
However, the scattering length of that interaction is positive, resulting
in a effective repulsion at low energy that prevents the Bose-Einstein
condensate from collapsing. One may therefore model the light boson
interaction as a repulsive interaction, while bearing in mind that
a realistic atomic interaction would in fact be attractive, and extra
bound states would exist below the range of energies where the interaction
has a repulsive effect, \emph{i.e.} below the dimer energy of two
light bosons.

\begin{figure}
\centering{}\includegraphics[scale=0.35]{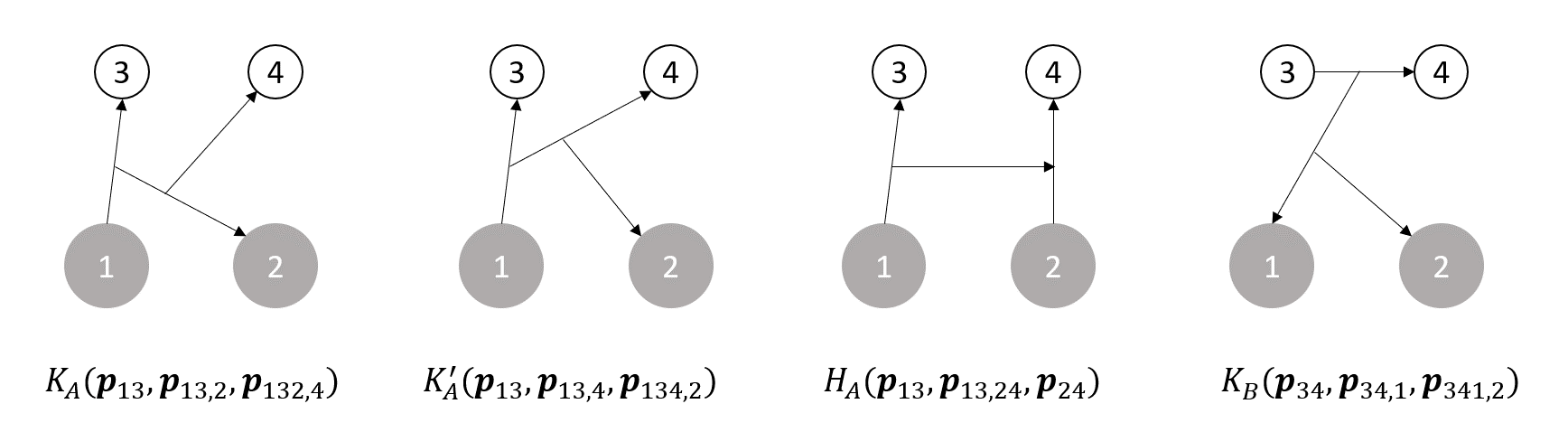}\caption{\label{fig:YakubovskyComponents}Schematic representation of the four
Yakubovsky components.}
\end{figure}

\section{System and method}

We consider the following four-body Hamiltonian,
\begin{equation}
\hat{H}=\frac{\hbar^{2}}{M}(\hat{\bm{p}}_{1}^{2}+\hat{\bm{p}}_{2}^{2})+\frac{\hbar^{2}}{m}(\hat{\bm{p}}_{3}^{2}+\hat{\bm{p}}_{4}^{2})+\hat{V}_{13}+\hat{V}_{14}+\hat{V}_{23}+\hat{V}_{24}+\hat{U}_{34},\label{eq:Hamiltonian}
\end{equation}
where $M$ is the mass of the two heavy particles, $m$ is the mass
of the two light particles, $\hat{V}$ is the heavy-light attractive
potential, and $\hat{U}$ is the light-light repulsive potential.
The two heavy particles are identical bosons called $A$, and the
two light particles are identical bosons called $B$. To simplify
the resolution, both potentials are taken to be separable, i.e. of
the forms
\begin{equation}
\hat{V}=\frac{4\pi\hbar^{2}}{2\mu}g\vert\phi\rangle\langle\phi\vert\label{eq:SeparableV}
\end{equation}
\begin{equation}
\hat{U}=\frac{4\pi\hbar^{2}}{m}\xi\vert\chi\rangle\langle\chi\vert\label{eq:SeparableU}
\end{equation}
where $\mu=(m^{-1}+M^{-1})^{-1}$ is the reduced mass, $g<0$ is an
attractive coupling strength and $\xi\ge0$ is a repulsive coupling
strength. After eliminating the centre of mass, the four-body Schrödinger
equation resulting from the Hamiltonian Eq.~(\ref{eq:Hamiltonian})
can be integrated as a set of integral equations on four-body components
known as the Yakubovsky equations~\cite{Yakubovsky1967,Gloeckle1993,Platter2004}.
For the most general problem of four distinguishable particles, there
are eighteen Yakubovsky components. In the present case, there are
only four, which are denoted as $K_{A}$, $K_{A}^{\prime},$ $H_{A}$,
and $K_{B}$, and represented schematically in Fig.~\ref{fig:YakubovskyComponents}.
These four components satisfy the four coupled integral equations:
\begin{align}
K_{A} & =\hat{G}_{0}\hat{T}_{13}\hat{P}_{12}(K_{A}+K_{A}^{\prime}+H_{A})\label{eq:Yakubovsky1}\\
K_{A}^{\prime} & =\hat{G}_{0}\hat{T}_{13}\left[\hat{P}_{34}(K_{A}+K_{A}^{\prime}+H_{A})+(1+\hat{P}_{12})K_{B}\right]\label{eq:Yakubovsky2}\\
H_{A} & =\hat{G}_{0}\hat{T}_{13}\left[\hat{P}_{12}\hat{P}_{34}(K_{A}+K_{A}^{\prime}+H_{A})\right]\label{eq:Yakubovsky3}\\
K_{B} & =\hat{G}_{0}\hat{T}_{34}\left[(1+\hat{P}_{34})(K_{A}+K_{A}^{\prime}+H_{A})\right]\label{eq:Yakubovsky4}
\end{align}
where $\hat{G}_{0}$ denotes the four-body Green's function operator
(without the centre of mass) at the four-body energy $E$, $\hat{P}_{ij}$
is the permutation operator for the coordinates of particles $i$
and $j$, and $\hat{T}_{ij}$ is the two-body $T$ matrix for the
two-body interaction between particles $i$ and $j$. In momentum
representation, we have:

\begin{equation}
G_{0}=\left(E-\frac{\hbar^{2}}{2\mu}p_{13}^{2}-\frac{\hbar^{2}}{2\mu_{13,2}}p_{13,2}^{2}-\frac{\hbar^{2}}{2\mu_{132,4}}p_{132,4}^{2}\right)^{-1},\label{eq:FourBodyGreenesFunction}
\end{equation}
where the wave vectors $\bm{p}_{13}$, $\bm{p}_{13,2}$, and $\bm{p}_{132,4}$
are the Fourier conjugates of the spatial vectors represented in the
leftmost picture of Fig.~\ref{fig:YakubovskyComponents}. For the
separable potentials Eqs.~(\ref{eq:SeparableV}) and (\ref{eq:SeparableU}),
we have:
\begin{equation}
\hat{T}_{13}=T_{13}(E_{13})\vert\phi\rangle_{13}\langle\phi\vert_{13}\qquad\text{ with }T_{13}(E)=\left(\frac{2\mu}{4\pi\hbar^{2}g}+\int\frac{d^{3}\bm{p}}{(2\pi)^{3}}\frac{\vert\phi(\bm{p})\vert^{2}}{\frac{\hbar^{2}p^{2}}{2\mu}-E}\right)^{-1}\label{eq:Tmatrix13}
\end{equation}
\begin{equation}
\hat{T}_{34}=T_{34}(E_{34})\vert\chi\rangle_{34}\langle\chi\vert_{34}\qquad\text{ with }T_{34}(E)=\left(\frac{m}{4\pi\hbar^{2}\xi}+\int\frac{d^{3}\bm{p}}{(2\pi)^{3}}\frac{\vert\chi(\bm{p})\vert^{2}}{\frac{\hbar^{2}p^{2}}{m}-E}\right)^{-1}\label{eq:Tmatrix34}
\end{equation}
where $E_{13}=E-\frac{\hbar^{2}}{2\mu_{13,2}}p_{13,2}^{2}-\frac{\hbar^{2}}{2\mu_{132,4}}p_{132,4}^{2}$
and $E_{34}=E-\frac{\hbar^{2}}{2\mu_{34,1}}p_{34,1}^{2}-\frac{\hbar^{2}}{2\mu_{34,12}}p_{34,12}^{2}$.
The indices of $\vert\phi\rangle$ of $\vert\chi\rangle$ denote the
relative coordinate of the pair on which they are acting. Thanks to
the separable form of the $T$ matrices Eq.~(\ref{eq:Tmatrix13})
and (\ref{eq:Tmatrix34}), the Yakubovsky equations can be written
as:
\begin{align}
K_{A} & =\hat{G}_{0}\vert\phi\rangle_{13}\bar{K}_{A}\label{eq:CompactYakubovsky1}\\
K_{A}^{\prime} & =\hat{G}_{0}\vert\phi\rangle_{13}\bar{K}_{A}^{\prime}\label{eq:CompactYakubovsky2}\\
H_{A} & =\hat{G}_{0}\vert\phi\rangle_{13}\bar{H}_{A}\label{eq:CompactYakubovsky3}\\
K_{B} & =\hat{G}_{0}\vert\chi\rangle_{34}\bar{K}_{B}\label{eq:CompactYakubovsky4}
\end{align}
where 
\begin{align}
\bar{K}_{A} & =T_{13}(E_{13})\langle\phi\vert_{13}P_{12}(K_{A}+K_{A}^{\prime}+H_{A})\label{eq:KAbar}\\
\bar{K}_{A}^{\prime} & =T_{13}(E_{13})\langle\phi\vert_{13}\left[\hat{P}_{34}(K_{A}+K_{A}^{\prime}+H_{A})+(1+\hat{P}_{12})K_{B}\right]\label{eq:KAPrimeBar}\\
\bar{H}_{A} & =T_{13}(E_{13})\langle\phi\vert_{13}\left[\hat{P}_{12}\hat{P}_{34}(K_{A}+K_{A}^{\prime}+H_{A})\right]\label{eq:HAbar}\\
\bar{K}_{B} & =T_{34}(E_{34})\langle\chi\vert_{34}\left[(1+\hat{P}_{34})(K_{A}+K_{A}^{\prime}+H_{A})\right]\label{eq:KBbar}
\end{align}
Inserting Eqs.~(\ref{eq:CompactYakubovsky1}-\ref{eq:CompactYakubovsky4})
into Eqs.~(\ref{eq:KAbar}-\ref{eq:KBbar}), one gets a closed set
of integral equations on $\bar{K}_{A}$, $\bar{K}_{A}^{\prime}$,
$\bar{H}_{A}$, and $\bar{K}_{B}$. Using the linear relations between
the various wave vectors of Fig.~\ref{fig:YakubovskyComponents},
one can write the final equations explicitly in momentum representation.
Note that these equations are easier to solve than the original Yakubovsky
equations (\ref{eq:Yakubovsky1}-\ref{eq:Yakubovsky4}) because the
number of (vector) arguments of $\bar{K}_{A}$, $\bar{K}_{A}^{\prime}$,
$\bar{H}_{A}$, and $\bar{K}_{B}$ is reduced by one from that of
$K_{A}$, $K_{A}^{\prime}$, $H_{A}$, and $K_{B}$, due to the integration
over one vector argument implied by the action of $\langle\phi\vert$
or $\langle\chi\vert$ in Eqs.~(\ref{eq:KAbar}-\ref{eq:KBbar}).
This reduction of arguments is the usual simplification coming from
the use of separable interactions. As a last simplification, we consider
the solutions with zero total angular momentum, and omit the angular
dependence of the vector arguments of $\bar{K}_{A}$, $\bar{K}_{A}^{\prime}$,
$\bar{H}_{A}$, and $\bar{K}_{B}$, which is known to be a very good
approximation, since the decomposition of the four-body wave function
in Yakubovsky components already captures most of its dependence on
the particles' angular momenta.

We are interested in expressing the results in terms of the scattering
lengths $a$ and $a_{B}$ of the potentials $\hat{V}$ and $\hat{U}$.
For this purpose, one can take $\hat{V}$ to the limit of a contact
interaction of scattering length $a$, by setting $\phi(p)=1$ up
to some arbitrarily large cutoff $\Lambda$, and $\phi(p)=0$ for
$p>\Lambda$, expressing the equations in terms of the scattering
length $a=(g^{-1}+\frac{2}{\pi}\Lambda)^{-1}$ and taking the limit
$\Lambda\to\infty$. Similarly, one can also set $\chi(p)=1$ for
$p<\Lambda_{B}$, and renormalise the result in terms of the scattering
length $a_{B}=(\xi^{-1}+\frac{2}{\pi}\Lambda_{B})^{-1}$. However,
the contact interaction limit $\Lambda_{B}\to\infty$ cannot be taken,
because $\xi$ being positive, it can only result in $a_{B}\to0$.
This is the well-known fact that a repulsive contact interaction in
three-dimensions does not scatter particles. On the opposite, one
can take $\xi\to\infty$, i.e. the strongest possible repulsion, which
results in $a_{B}=\frac{\pi}{2}\Lambda_{B}^{-1}$. With these choices,
$\hat{V}$ is only parameterised by $a$, and $\hat{U}$ is only parameterised
by $a_{B}\ge0$.

Because of the Efimov effect occuring for two heavy and one light
bosons as well as one heavy and two light bosons, such three particles
can experience an Efimov attraction from large distances to distances
comparable to the range of $\hat{V}$. Since here the range of $\hat{V}$
is taken to zero by setting $\Lambda\to\infty$, nothing prevents
the three particles from collapsing to a single point, which makes
the problem ill-defined with no ground state. For such zero-range
interactions, an extra three-body parameter must be introduced to
cure this problem and set the energy of the trimers to a finite value.
Here, it is implemented by imposing a finite cutoff $\Lambda_{3}$
on the wave vectors $\bm{p}_{13,2}$, $\bm{p}_{14,2}$, $\bm{p}_{23,1}$
and $\bm{p}_{24,1}$ between a heavy-light pair and a heavy particle,
as well as the wave vectors $\bm{p}_{31,4}$, $\bm{p}_{32,4}$, $\bm{p}_{41,3}$
and $\bm{p}_{42,3}$ between a heavy-light pair and a light particle.
In the end, the three parameters of the system are the two scattering
lengths $a$ and $a_{B}$, and the length $\Lambda_{3}^{-1}$ which
corresponds to the range of interactions, on the order of a few nanometres
for atoms.

\begin{figure}
\includegraphics[width=17cm]{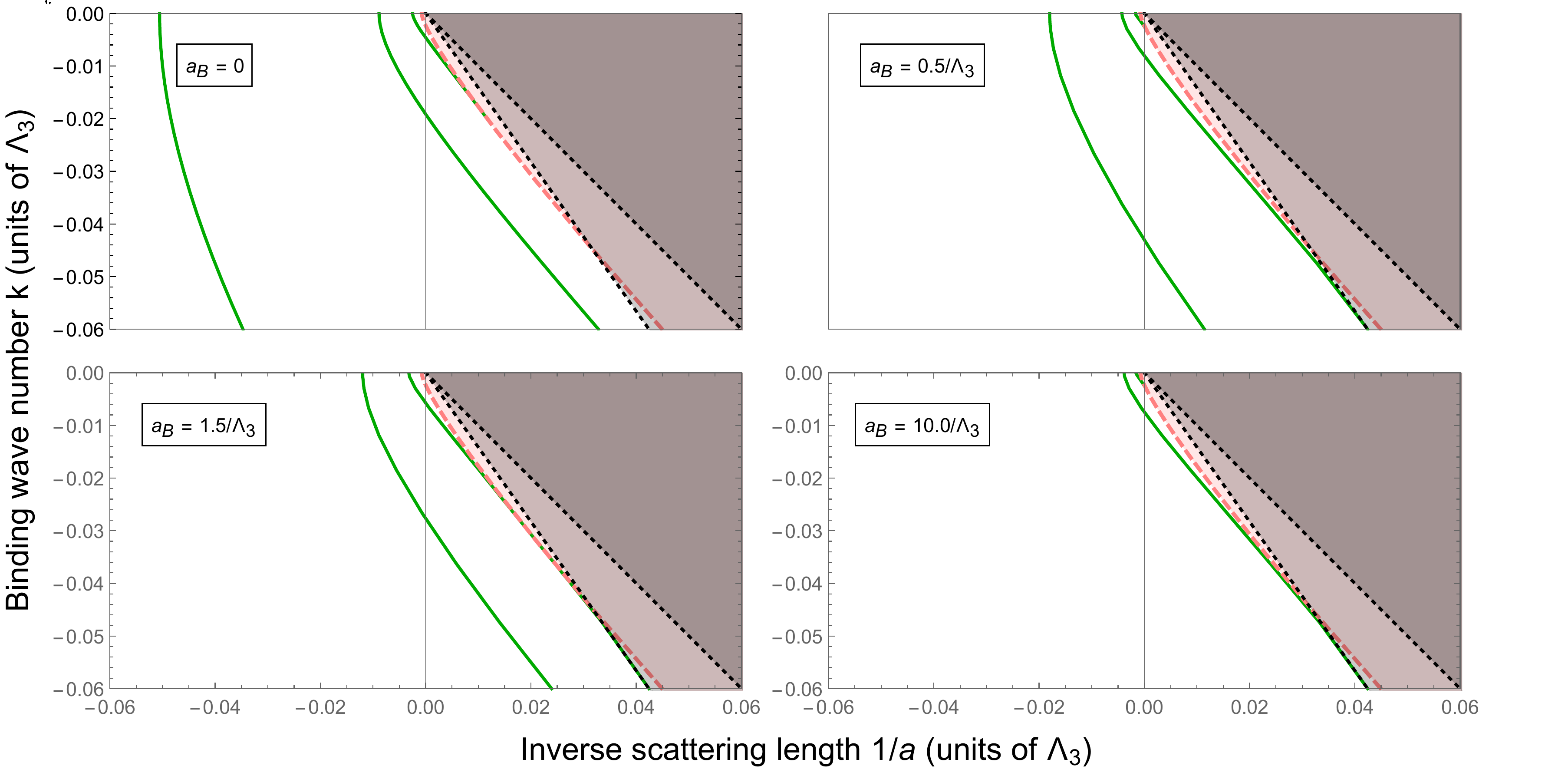}

\caption{\label{fig:Equal-mass}Energy spectrum (rescaled as a binding wave
number $k=-\sqrt{2\mu\vert E\vert}/\hbar$) for a system of two identical
bosons $A$ and two identical bosons $B$ of equal mass, as a function
of the inverse scattering length $1/a$ between the two kinds of bosons.
Each panel corresponds to a different value of the scattering length
$a_{B}$ between two bosons $B$, as indicated at the top of each
panel. The black dotted lines represent the energies of one and two
$AB$ dimers. The red dashed curve represents the energy of the $AAB$
ground trimer. These curves correspond to the thresholds of scattering
continua indicated by shaded areas. The green solid curves represent
the $AABB$ tetramer bound states.}
\end{figure}

\section{Results}

In the following calculations, the range $\Lambda_{3}^{-1}$ is taken
as the unit of length and $E_{3}=\hbar^{2}\Lambda_{3}^{2}/(2\mu)$
as the unit of energy. One can then fix $a_{B}$ and vary $1/a$ from
negative to positive values, corresponding to increasing attractive
strength $g$ between heavy and light particles. The point $1/a=0$
corresponds to the unitary point at which a heavy-light $AB$ dimer
is formed, and around which the Efimov effect occurs for $AAB$ and
$ABB$ systems.

\begin{figure}
\includegraphics[width=17cm]{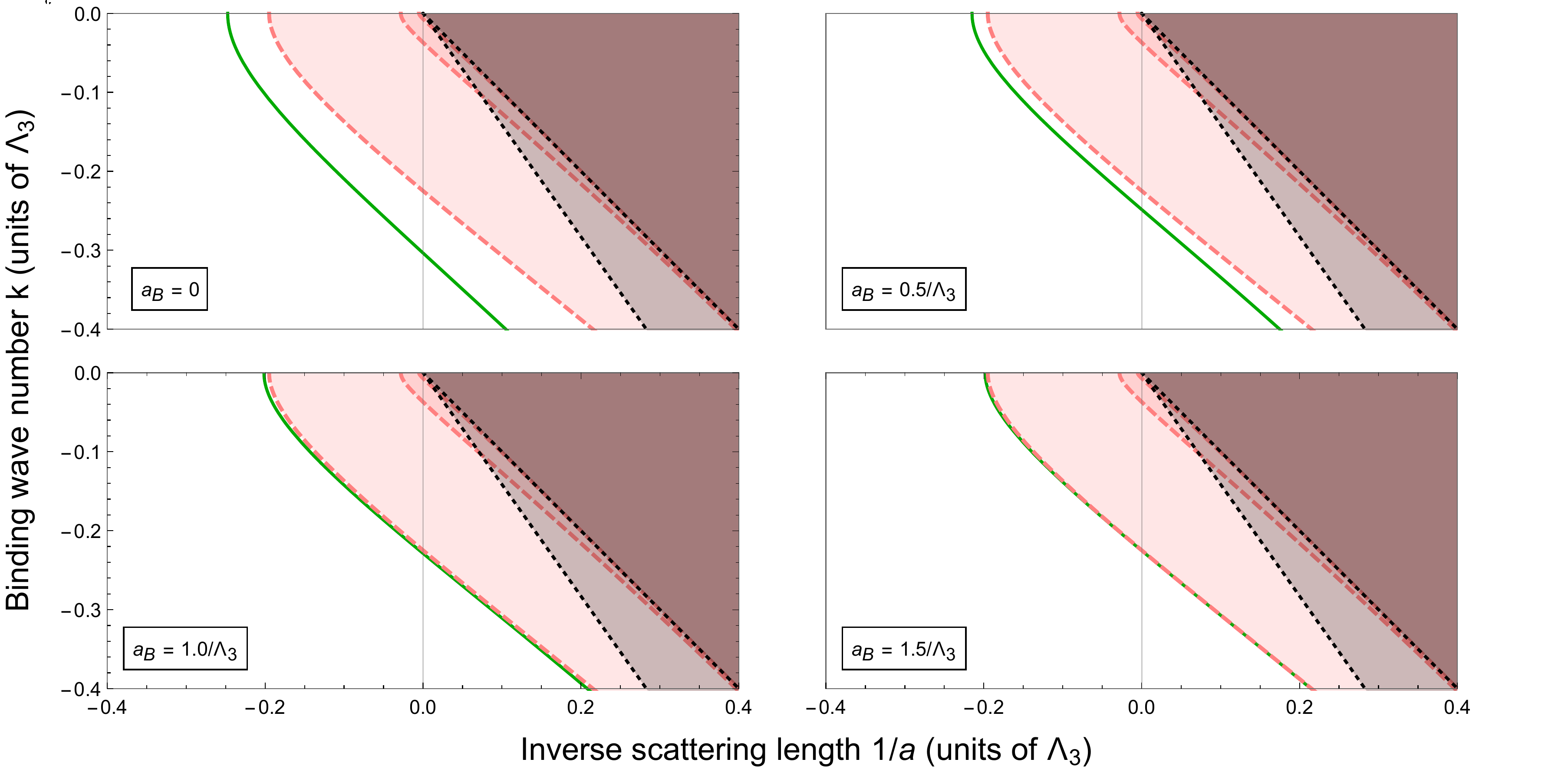}\caption{\label{fig:MassRatio19}Same as figure 2 for a system of two heavy
bosons $A$ and two light bosons $B$, with mass ratio 19. Each panel
corresponds to a different value of the scattering length $a_{B}$
between the light bosons $B$. }
\end{figure}

\subsection{Equal-mass case}

Let us first consider the case of equal masses $M=m$. In this case,
the Efimov attraction is weak, with a scaling factor $e^{\pi/\vert s_{0}\vert}\approx1986$
between consecutive trimer states~\cite{Naidon2017}. In the energy
spectrum as function of $1/a$ shown in Fig.~\ref{fig:Equal-mass},
only the ground-state $AAB$ trimer (degenerate with the ground-state
$ABB$ trimer) can be seen, the other trimer states being too weakly
bound to be seen. For $a_{B}=0$, corresponding to no interaction
between the $B$ particles, one finds at least three Borromean $AABB$
tetramer states below the ground-state trimer (there might be more
weakly bound tetramers that are not resolved in the present numerical
calculation). The tetramers are here significantly more bound than
the trimers: at unitarity ($1/a=0$), the ground-state tetramer binding
energy $0.016E_{3}$ is 400 times larger than the ground-state trimer
binding energy $0.000037E_{3}$.

As the value of $a_{B}$ is increased, the tetramers are weakened
and gradually pushed to the $AAB+B$ threshold, as seen in Fig.~\ref{fig:Equal-mass}.
Although they are significantly weakened, it is not possible to fully
unbind all the tetramers, even for $a_{B}$ as large as $10\Lambda_{3}^{-1}$.
For larger values of $a_{B}$, the tetramers become more strongly
bound again, and in the limit $a_{B}\to\infty$, one retrieves the
original spectrum obtained for $a_{B}=0$. This is because the repulsive
effect of the boson interaction becomes more and more diluted over
large distances as $a_{B}$ is increased. In the limit of large $a_{B}$,
only the weak tetramers with a size on the order of $a_{B}$ are signicantly
affected by the light boson repulsion. This situation qualitatively
reproduces what would happen with an attractive $BB$ interaction
with large positive scattering length $a_{B}$: only the tetramers
with an energy above the $BB$ dimer energy $-\hbar^{2}/(ma_{B}^{2})$
would be pushed up, while the tetramers below that energy would be
pushed down, due to the avoided crossing with the $BB$ dimer. However,
the results are not expected to be quantitative for such large values
of $a_{B}$, so we shall disregard this regime and focus on $a_{B}\lesssim\Lambda_{3}$. 

It should be noted that if the boson-boson interaction were modelled
by a hard-core repulsion, all tetramers would of course unbind for
$a_{B}\gtrsim10\Lambda_{3}$, although this would not be a physically
correct description of atoms with large scattering length.

\begin{figure}
\begin{centering}
\includegraphics[height=7cm]{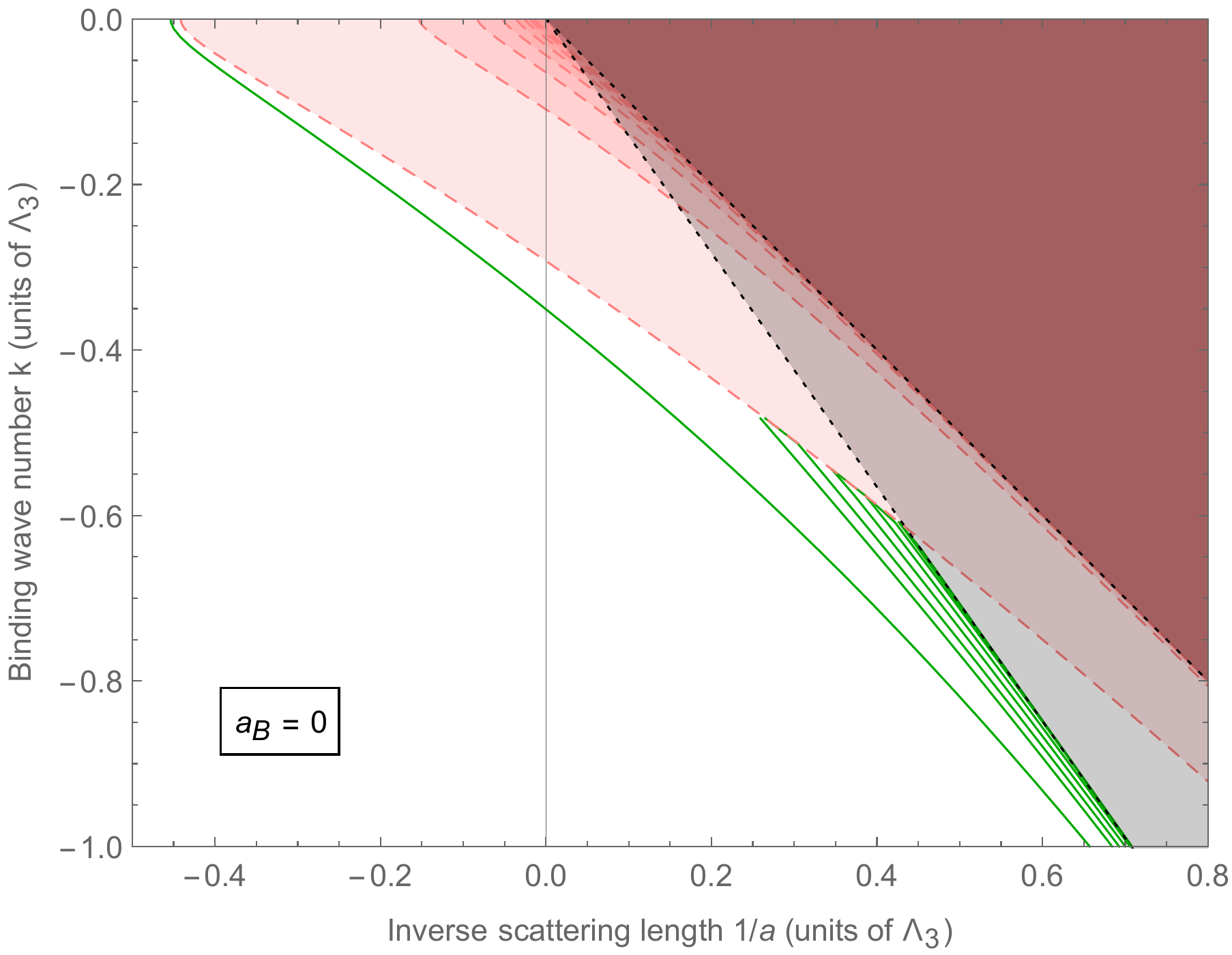}\includegraphics[bb=55bp 0bp 520bp 403bp,clip,height=7cm]{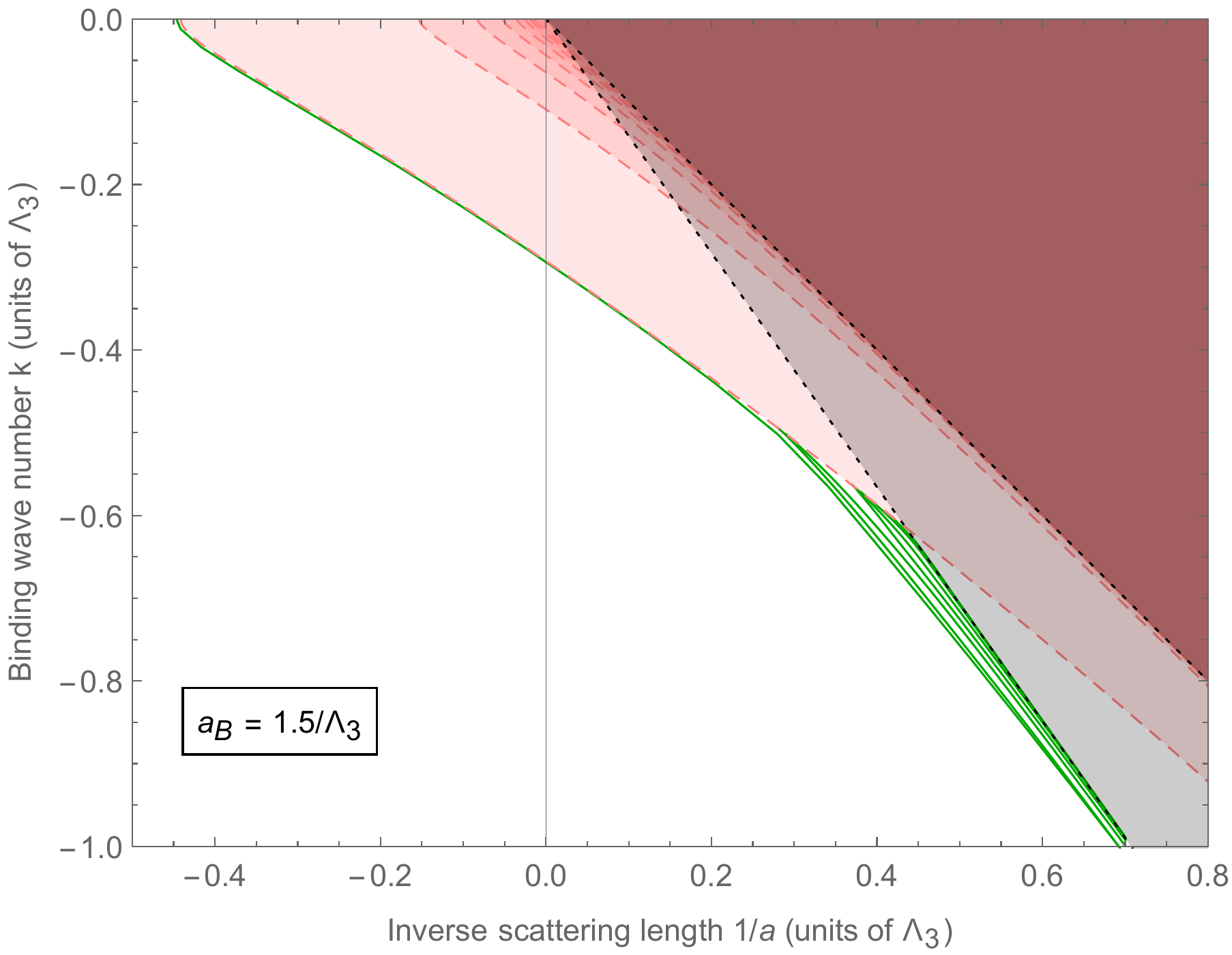}
\par\end{centering}
\caption{\label{fig:MassRatio1000}Same as figure 2 for a system of two heavy
bosons $A$ and two light bosons $B$, with mass ratio 1000. Each
panel corresponds to a different value of the scattering length $a_{B}$
between the light bosons $B$. }

\end{figure}

\begin{figure}
\begin{centering}
\includegraphics[height=7cm]{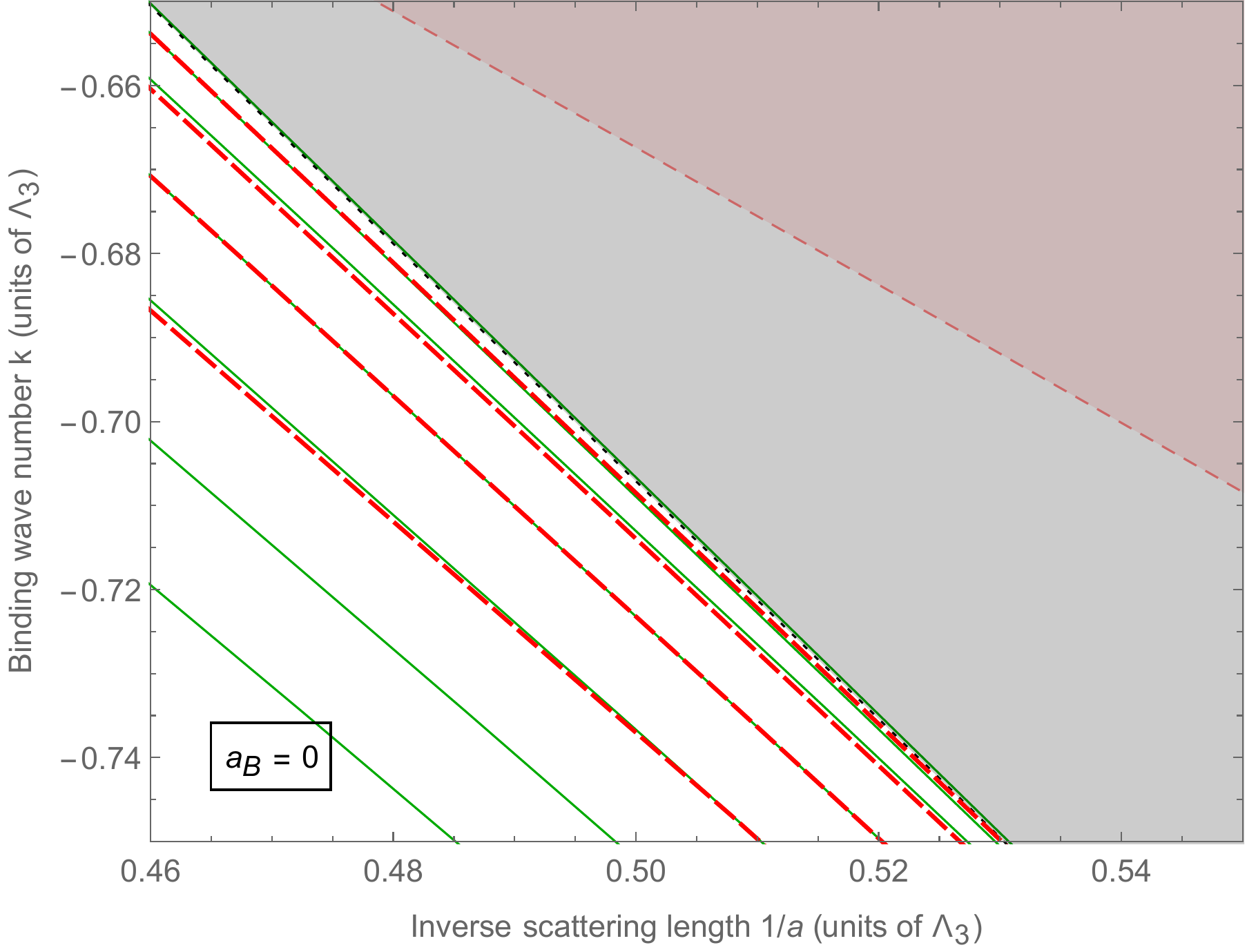}\caption{\label{fig:MassRatioZoom}Close up of the left panel of Fig.~4. The
red long-dashed curves correspond to the 5th tetramer curve scaled
by the coefficient $\lambda^{n},$ for $n=-1,0,1,2$, where $\lambda=1.1913$
is obtained from Eq.~(\ref{eq:TetramerScalingFactor}).}
\par\end{centering}
\end{figure}

\subsection{Large mass imbalance $M/m=19$}

For larger mass imbalance, the Efimov attraction mediated by a particle
$B$ is stronger and the $AAB$ trimers are more strongly bound, whereas
the Efimov attraction mediated by a particle $A$ is weaker and the
$ABB$ trimers are more weakly bound. We consider the mass ratio $M/m=19$
corresponding to a mixture of two cesium-133 and two lithium-7 atoms.
In the energy spectrum shown in Fig.~\ref{fig:MassRatio19}, the
lowest three $AAB$ trimers can be seen, while the $ABB$ trimers
are too weak to be seen. The ground-state $AAB$ trimer is now bound
with a binding energy of $0.053E_{3}$ at unitarity. For $1/a<0$
and around unitarity, one finds only one Borromean $AABB$ tetramer
state for $a_{B}=0$. Its binding energy at unitarity is $0.092E_{3}$.
While both the trimers and tetramers are more strongly bound with
respect to the equal-mass case, their relative separation is reduced.
As a result, it is easier to push the tetramer to the $AAB+B$ threshold.
Fig.~\ref{fig:MassRatio19} shows that the tetramer nearly reaches
the $AAB+B$ threshold for $a_{B}\gtrsim1.5\Lambda_{3}^{-1}$. However,
as in the equal-mass case, it is not possible to fully unbind the
tetramer, and for larger values of $a_{B}$ the tetramer gradually
recovers its original binding energy.

\subsection{Very large mass imbalance $M/m=1000$}

Although the mass ratio $M/m=19$ may be considered as large, only
one tetramer bound state is found near unitarity. As mentioned in
the introduction, at large mass ratio, the two heavy particles are
expected to undergo twice the Efimov attraction $V(R)=-(\vert s_{0}\vert^{2}+1/4)/R^{2}$
mediated by each of the two light bosons, resulting in an infinite
Efimov series of $AABB$ tetramers. The scaling factor of this Efimov
series is therefore
\begin{equation}
\lambda=e^{\pi/s_{0}^{\prime}},\qquad\text{ with }s_{0}^{\prime}=\sqrt{2\vert s_{0}\vert^{2}+1/4},\label{eq:TetramerScalingFactor}
\end{equation}
where $\vert s_{0}\vert$ is given by Eq.~(\ref{eq:s0}).

It is however not easy to observe these tetramers at mass ratio 19,
because they are hidden as resonant states in the $AAB+B$ scattering
continuum, which are difficult to extract within the present calculations.
Nevertheless, by further increasing the mass ratio, one can observe
more tetramer bound states emerging from the $AAB+B$ threshold, as
shown in Fig.~\ref{fig:MassRatio1000} for the mass ratio $M/m=1000$.
Figure~\ref{fig:MassRatioZoom} shows that these tetramers form indeed
a discrete scale invariant spectrum with the scaling factor $\lambda$
given in Eq.~(\ref{eq:TetramerScalingFactor}). Although they are
expected to persists inside the $AAB+B$ scattering continuum as universal
four-body resonances, their Efimov spectrum may be significantly altered
by its multiple crossings with the excited trimer+particle thresholds.

Finally, it can be checked again that the ground-state tetramer can
be pushed near dissociation into a $AAB$ trimer and $B$ particle
for a moderately repulsive interaction between the light bosons $B$.
Such a situation is shown in the right panel of Fig.~\ref{fig:MassRatio1000}
for $a_{B}=1.5\Lambda_{3}^{-1}$ .

\section{Conclusion}

In a system of two heavy bosons and two light bosons, a near-resonant
($a\gg\Lambda_{3}^{-1}$) attractive force between the heavy and light
bosons is required to overcome quantum fluctuations and bind the system
into tetramers. On the other hand, a relatively weak non-resonant
$(a_{B}\sim\Lambda_{3}^{-1})$ repulsion between the light bosons
can nearly unbind the tetramers. This suggests that in the polaron
problem of heavy bosonic impurities immersed in a Bose-Einstein condensate
of light bosons, the attraction between two impurities mediated by
the condensate is carried mostly by a single light boson (or single
excitation) for a relatively moderate repulsion between the light
bosons.

\bibliographystyle{IEEEtranS}

\end{document}